\documentclass[12pt,a4paper]{article}
\usepackage{graphicx}
\usepackage{amsmath}
\usepackage{amsfonts}
\usepackage{amssymb}

\begin{document}

\title{\textbf{Integrability of the Higher-Order}\\\textbf{Nonlinear Schr\"{o}dinger Equation}\\\textbf{Revisited}}
\author{\textsc{S. Yu. Sakovich\bigskip}\\\textit{Institute of Physics,}\\\textit{National Academy of Sciences,}\\\textit{P.O.72, Minsk, Belarus}\\\textit{sakovich@dragon.bas-net.by}}
\date{\textit{June 22, 1999}}
\maketitle
\begin{abstract}
Only the known integrable cases of the Kodama-Hasegawa higher-order nonlinear
Schr\"{o}dinger equation pass the Painlev\'{e} test. Recent results of Ghosh
and Nandy add no new integrable cases of this equation.\bigskip
\end{abstract}

Being a true high-technology application of a mathematical object, the optical
soliton in fiber optical lines of communications is an ideal carrier of
information because the integrability\emph{ }of its model equation, the
nonlinear Schr\"{o}dinger (NLS) equation, provides a guarantee on the input
and output causal relation for light waves in fibers \cite{Has}. The
integrable NLS equation, however, does not govern femtosecond light pulses
which have much potential for the future technology. Kodama and Hasegawa
\cite{KH} derived a model equation for ultra-short light pulses in optical
fibers, the higher-order nonlinear Schr\"{o}dinger (HNLS) equation:
\begin{equation}
iw_{z}+\frac{1}{2}w_{yy}+\left|  w\right|  ^{2}w+i\alpha w_{yyy}+i\beta\left|
w\right|  ^{2}w_{y}+i\gamma w(\left|  w\right|  ^{2})_{y}=0,\label{hnls}%
\end{equation}
where real parameters $\alpha$, $\beta$ and $\gamma$ are determined by
spectral and geometric properties of a fiber. Eq. (\ref{hnls}) admits
one-soliton solutions of bright and dark types in wide domains of its
parameters \cite{PT}, but this has no relation to its integrability. Only the
following four integrable\emph{ }cases of the HNLS eq. (\ref{hnls}) are known
besides the NLS equation itself: the derivative NLS equation I \cite{KN}
($\alpha:\beta:\gamma=0:1:1$), the derivative NLS equation II \cite{CLL}
($0:1:0$), the Hirota equation \cite{Hir} ($1:6:0$), and the Sasa-Satsuma
equation \cite{SS} ($1:6:3$). According to the results of Nijhof and Roelofs
\cite{NR}, only these known integrable cases of the HNLS eq. (\ref{hnls})
possess the infinite-dimensional prolongation structures.

The Painlev\'{e} analysis leads us to the same conclusion on the integrability
of eq. (\ref{hnls}). Let us remind in brief our results obtained in \cite{Sak}
in the framework of the Weiss-Kruskal algorithm (for details of the method see
e.g. \cite{RGB} and references therein).

The HNLS eq. (\ref{hnls}) with $\alpha=0$ lies in the class of derivative NLS
equations which has been analyzed by Clarkson and Cosgrove \cite{CC}. We find
from their results \cite{CC} that eq. (\ref{hnls}) with $\alpha=0$ has the
Painlev\'{e} property if and only if $(\beta-\gamma)\gamma=0$, i.e. exactly
when eq. (\ref{hnls}) is the derivative NLS equations I and II besides the NLS
equation itself.

When $\alpha\neq0$, we can transform eq. (\ref{hnls}) by
\begin{align}
w(y,z) &  =u(x,t)\exp\left(  \frac{i}{6}\alpha^{-1}x-\frac{i}{216}\alpha
^{-3}t\right)  ,\nonumber\\
y &  =x-\frac{1}{12}\alpha^{-2}t,\text{\qquad}z=-\alpha^{-1}t\label{trans}%
\end{align}
into the equivalent complex modified Korteweg-de Vries (CMKdV) equation
\begin{equation}
u_{t}=u_{xxx}+auu^{\ast}u_{x}+bu^{2}u_{x}^{\ast}+icu^{2}u^{\ast},\label{cmkdv}%
\end{equation}
where * denotes the complex conjugation, $a=\alpha^{-1}(\beta+\gamma)$,
$b=\alpha^{-1}\gamma$, and $c=\frac{1}{6}\alpha^{-2}(\beta-6\alpha)$. (It is
sometimes overlooked in the literature that the CMKdV eq. (\ref{cmkdv}) with
$c=0$ is equivalent to the HNLS eq. (\ref{hnls}) only if $\beta=6\alpha$ in
eq. (\ref{hnls}).) Since eq. (\ref{cmkdv}) is a complex equation, we
supplement eq. (\ref{cmkdv}) by its complex conjugation, introduce the new
variable $v$, $v=u^{\ast}$, and then consider $u$ and $v$ as mutually
independent. Thus, we have the following system of two nonlinear equations of
total order six:
\begin{align}
u_{xxx}+auvu_{x}+bu^{2}v_{x}+icu^{2}v-u_{t} &  =0,\nonumber\\
v_{xxx}+auvv_{x}+bv^{2}u_{x}-icuv^{2}-v_{t} &  =0.\label{sys}%
\end{align}

A hypersurface $\varphi(x,t)=0$ is non-characteristic for this system if
$\varphi_{x}\neq0$ (we take $\varphi_{x}=1$), and the general solution of eq.
(\ref{sys}) must contain six arbitrary functions of one variable. We
substitute the expressions $u=\varphi^{\sigma}[u_{0}(t)+...+u_{r}\left(
t\right)  \varphi^{r}+...]$ and $v=\varphi^{\tau}[v_{0}(t)+...+v_{r}\left(
t\right)  \varphi^{r}+...]$ into eq. (\ref{sys}) for determining the exponents
$\sigma$ and $\tau$ of the dominant behavior of solutions and the positions
$r$ of resonances, and find the following two branches if $a^{2}+b^{2}\neq0$
(we reject the case $a=b=0$, $c\neq0$ because of inadmissible $r=\frac{1}%
{2}\left(  5\pm i\sqrt{87}\right)  $ ).

\begin{itemize}
\item \textbf{Branch (A): }$\sigma=\tau=-1$, $u_{0}v_{0}=-6\left(  a+b\right)
^{-1}$, $u_{0}/v_{0}$ is arbitrary, $r=-1,0,3,4,3-p,3+p$, where
\begin{equation}
p=2\left(  \frac{a-2b}{a+b}\right)  ^{1/2}. \label{p}%
\end{equation}

\item \textbf{Branch (B): }$\sigma=-1\pm n$, $\tau=-1\mp n$, $u_{0}%
v_{0}=-60\left(  5a-7b\right)  ^{-1}$, $u_{0}/v_{0}$ is arbitrary,
$r=-1,0,4,6,\frac{1}{2}\left(  3-q\right)  ,\frac{1}{2}\left(  3+q\right)  $,
where
\begin{equation}
n=\left(  \frac{5a+17b}{5a-7b}\right)  ^{1/2}, \label{n}%
\end{equation}%
\begin{equation}
q=\left(  \frac{245a+617b}{5a-7b}\right)  ^{1/2}. \label{q}%
\end{equation}
\end{itemize}

We reject the following cases: $a+b=0$, when the branch (A) does not exist,
because of inadmissible $r=\frac{1}{2}\left(  3\pm i\sqrt{31}\right)  $ in the
branch (B); $5a-7b=0$, when the branch (B) does not exist, because of
inadmissible $r=3\pm i$ in the branch (A); and $5a+17b=0$, when the branches
(A) and (B) coincide, because the double resonance $r=0$ and the fact that
$u_{0}v_{0}$ is determined require logarithmic terms in the singular
expansions. Thus, the two different branches, (A) and (B), exist for all the
cases of eq. (\ref{cmkdv}) we have to analyze farther. The existence of the
branch (B) was ignored in several works, where some special cases of the CMKdV
\emph{\ }eq. (\ref{cmkdv}) were tested. The branch (B) was lost as well in
\cite{PN}, where the Painlev\'{e} test of the HNLS eq. (\ref{hnls}) missed all
the integrable cases except the Sasa-Satsuma case.

Eqs. (\ref{p}), (\ref{n}) and (\ref{q}) show that the dominant behavior of
solutions in the branch (B) and the positions of two resonances in each of the
branches (A) and (B) are determined only by the quotient $a/b$. Elimination of
$a/b$ from eqs. (\ref{p}), (\ref{n}) and (\ref{q}) leads to the following two
equations:
\begin{equation}
\left(  1+p^{2}\right)  \left(  1+n^{2}\right)  =10,\label{frst}%
\end{equation}%
\begin{equation}
9+40n^{2}=q^{2}.\label{scnd}%
\end{equation}
The numbers $p$, $n$ and $q$ have to be integer for the CMKdV \emph{\ }eq.
(\ref{cmkdv}) to possess the Painlev\'{e} property. Eqs. (\ref{frst}) and
(\ref{scnd}) admit three integer solutions: $(p,n,q)=(1,2,13),(2,1,7),(3,0,3)$%
, but the last one corresponds to the already rejected case $a/b=-17/5$. The
solution $(2,1,7)$ leads to the Hirota case of the HNLS eq. (\ref{hnls}):
$b=0$ in eq. (\ref{cmkdv}) corresponds to $\gamma=0$ in eq. (\ref{hnls}), the
usual way of constructing recursion relations and checking compatibility
conditions at resonances gives us the condition $c=0$ (i.e. $\beta=6\alpha$ in
eq. (\ref{hnls})) at $r=1$ in the branch (A), and then compatibility
conditions become identities in both branches. The solution $(1,2,13)$ leads
to the Sasa-Satsuma case of the HNLS eq. (\ref{hnls}): $\beta=2\gamma$ since
$a=3b$, the condition $c=0$ (i.e. $\beta=6\alpha$) arises at $r=3$ in the
branch (A), and other compatibility conditions are all satisfied.

Consequently, only the known integrable cases of the HNLS eq. (\ref{hnls})
pass the Painlev\'{e} test for integrability \cite{Sak}. This completely
agrees with the results of Nijhof and Roelofs \cite{NR}.\bigskip

Recently, however, Ghosh and Nandy \cite{GN} reported that they found a
parametric Lax representation for the CMKdV eq. (\ref{cmkdv}) with $c=0$ and
any rational value of $b/a$ from the interval $[0,1]$. Let us consider in
brief the intriguing results of \cite{GN}.

The Lax pair, proposed in \cite{GN}, is as follows:%
\begin{equation}
\Psi_{x}=U\Psi,\qquad\Psi_{t}=V\Psi, \label{lax}%
\end{equation}%
\begin{equation}
U=-i\lambda\Sigma+A, \label{mu}%
\end{equation}%
\begin{equation}
V=A_{xx}+AA_{x}-A_{x}A-2A^{3}-2i\lambda\Sigma(A_{x}-A^{2})-4\lambda
^{2}A+4i\lambda^{3}\Sigma, \label{mv}%
\end{equation}%
\begin{equation}
\Sigma=\left(
\begin{array}
[c]{ccccccc}%
1 & \cdots & 0 & 0 & \cdots & 0 & 0\\
\vdots & \ddots & \vdots & \vdots & \ddots & \vdots & \vdots\\
0 & \cdots & 1 & 0 & \cdots & 0 & 0\\
0 & \cdots & 0 & 1 & \cdots & 0 & 0\\
\vdots & \ddots & \vdots & \vdots & \ddots & \vdots & \vdots\\
0 & \cdots & 0 & 0 & \cdots & 1 & 0\\
0 & \cdots & 0 & 0 & \cdots & 0 & -1
\end{array}
\right)  , \label{ms}%
\end{equation}%
\begin{equation}
A=\left(
\begin{array}
[c]{ccccccc}%
0 & \cdots & 0 & 0 & \cdots & 0 & u\\
\vdots & \ddots & \vdots & \vdots & \ddots & \vdots & \vdots\\
0 & \cdots & 0 & 0 & \cdots & 0 & u\\
0 & \cdots & 0 & 0 & \cdots & 0 & u^{\ast}\\
\vdots & \ddots & \vdots & \vdots & \ddots & \vdots & \vdots\\
0 & \cdots & 0 & 0 & \cdots & 0 & u^{\ast}\\
-u^{\ast} & \cdots & -u^{\ast} & -u & \cdots & -u & 0
\end{array}
\right)  , \label{ma}%
\end{equation}
where $\lambda$ is a parameter, $u=u(x,t)$, and the dimension of the
block-form matrices $\Sigma$ and $A$ is $(l+m+1)\times(l+m+1)$ (note: the
expression for $V$ contains some misprints in \cite{GN}). This form of $U$ and
$V$ provides that the compatibility condition of the system (\ref{lax}),
$U_{t}=V_{x}-UV+VU$, is as follows:%
\begin{equation}
A_{t}=A_{xxx}-3(A^{2}A_{x}+A_{x}A^{2}). \label{eqa}%
\end{equation}

It is stated in \cite{GN} that the equation (\ref{eqa}) with $A$ (\ref{ma}) is
nothing but%
\begin{equation}
u_{t}=u_{xxx}+(6l+3m)uu^{\ast}u_{x}+3mu^{2}u_{x}^{\ast} \label{gn}%
\end{equation}
and therefore the system (\ref{lax}) with (\ref{mu}), (\ref{mv}), (\ref{ms})
and (\ref{ma}) represents a Lax pair for eq. (\ref{cmkdv}) with $c=0$ and a
rational value of $b/a$ determined by the dimension of the block matrices.

The real situation, however, is different. There are three distinct cases of
what is in fact the equation (\ref{eqa}) with $A$ (\ref{ma}), depending on
values of $l$ and $m$.

\begin{itemize}
\item  If $m=0,l\neq0$ or $l=0,m\neq0$, then eq. (\ref{eqa}) is eq.
(\ref{cmkdv}) with $b=0$. This is the Hirota case.

\item  If $m=l\neq0$, then eq. (\ref{eqa}) is eq. (\ref{cmkdv}) with $a=3b$.
This is the Sasa-Satsuma case.

\item  If $m\neq0,l\neq0,m\neq l$, then eq. (\ref{eqa}) is the following
over-determined system of two complex evolution equations:%
\begin{align}
u_{t}  &  =u_{xxx}+(6l+3m)uu^{\ast}u_{x}+3mu^{2}u_{x}^{\ast},\nonumber\\
u_{t}  &  =u_{xxx}+(3l+6m)uu^{\ast}u_{x}+3lu^{2}u_{x}^{\ast}. \label{over}%
\end{align}
Representing $u$ as $u=fe^{ig}$, where $f$ and $g$ are real functions of $x$
and $t$, we get the following equivalent form of the system (\ref{over}):%
\begin{equation}
f_{t}=f_{xxx}+6(l+m)f^{2}f_{x},\qquad g_{x}=g_{t}=0. \label{mkdv}%
\end{equation}
Though integrable, the system (\ref{mkdv}) is only a reduction of the CMKdV
eq. (\ref{cmkdv}).
\end{itemize}

Consequently, no new integrable cases of the Kodama-Hasegawa HNLS eq.
(\ref{hnls}) were found in \cite{GN}.\bigskip

This work was supported in part by Grant $\Phi$98-044 of the Fundamental
Research Fund of Belarus.

\end{document}